\documentclass[10pt,a4paper]{article}
\usepackage[english]{babel}
\usepackage[utf8]{inputenc}
\usepackage{amsfonts,amsbsy,bm,euscript,mathrsfs}
\usepackage{amssymb,stmaryrd,faktor}
\usepackage[tbtags]{amsmath}
\usepackage{bbm}
\usepackage{graphicx}
\usepackage[title,titletoc]{appendix}
\usepackage[bookmarks=true,colorlinks=true,linkcolor=blue,citecolor=blue,urlcolor=blue,bookmarksnumbered]{hyperref}

\usepackage{dsfont}
\usepackage{collref}
\usepackage{lmodern}
\usepackage{mathrsfs}
\usepackage{mathtools}
\usepackage{bbm}
\usepackage{braket}
\usepackage{slashed}

\usepackage{tikz-cd}
\usepackage{arydshln}
\tikzset{
    commutative diagrams/.cd,
    arrow style=tikz,
    diagrams={>={Computer Modern Rightarrow[length=5pt,width=5pt]}},
}
\usepackage{ulem}

\usepackage{graphicx}
\usepackage{booktabs}
\usepackage{subfig}
\usepackage{tikz}
\usetikzlibrary{plotmarks,calc,decorations,decorations.pathmorphing}

\numberwithin{equation}{section}

\makeatletter
\renewcommand\section{\@startsection {section}{1}{\z@}
{-3.5ex \@plus -1ex \@minus -.2ex}
{2.3ex \@plus.2ex}
{\normalfont\Large\bfseries}}
\renewcommand\subsection{\@startsection{subsection}{2}{\z@}
{-3.25ex\@plus -1ex \@minus -.2ex}
{1.5ex \@plus.2ex}
{\normalfont\large\bfseries}}
\makeatother

\newcommand{\alg}[1]{\mathfrak{#1}}

\def\smallL{\small{L}}
\def\smallR{\small{R}}

\newcommand{\bea}{\begin{eqnarray}}
\newcommand{\eea}{\end{eqnarray}}

\allowdisplaybreaks 

\DeclareMathOperator{\arccot}{arccot}

\usepackage{blkarray}

\begin{document}

\thispagestyle{empty}
\begin{flushright}\footnotesize\ttfamily
DMUS-MP-20/08
\end{flushright}
\vspace{2em}

\begin{center}

{\Large\bf \vspace{0.2cm}
{\color{black} \large Boosts superalgebras based on centrally-extended $\alg{su}(1|1)^2$}} 
\vspace{1.5cm}

\textrm{Juan Miguel Nieto García\footnote{\texttt{j.nietogarcia@surrey.ac.uk}}, Alessandro Torrielli \footnote{\texttt{a.torrielli@surrey.ac.uk}} and Leander Wyss\footnote{\texttt{l.wyss@surrey.ac.uk}}}

\vspace{2em}

\vspace{1em}
\begingroup\itshape
Department of Mathematics, University of Surrey, Guildford, GU2 7XH, UK
\par\endgroup

\end{center}

\vspace{2em}

\begin{abstract}\noindent 
In this paper, we studied the boost operator in the setting of $\mathfrak{su}(1|1)^2$. We find a family of different algebras where such an operator can consistently appear, which we classify according to how the two copies of the $\mathfrak{su}(1|1)$ interact with each other. Finally, we construct coproduct maps for each of these algebras and discuss the algebraic relationships among them.
\end{abstract}

\newpage

\overfullrule=0pt
\parskip=2pt
\parindent=12pt
\headheight=0.0in \headsep=0.0in \topmargin=0.0in \oddsidemargin=0in

\vspace{-3cm}
\thispagestyle{empty}
\vspace{-1cm}

\tableofcontents

\setcounter{footnote}{0}

\section{\label{sec:Intro}Introduction}

The infinite-dimensional quantum supergroup underlying the integrable systems beneath the $AdS_5/CFT_4$ correspondence \cite{Beisert:2010jr,Arutyunov:2009ga} is a constant source of new developments which extend our understanding of exact $S$-matrices and their (super)symmetries.   

Such an exotic Hopf superalgebra \cite{Janik:2006dc,B1,B2} is extraordinarily close to a Yangian \cite{Drinfeld:1985rx,Molev,BYa} (see also \cite{Dsec}), but it is distinct from it in ways that allow the possibility of in fact a much larger and more complicated structure. One can certainly recognise in it a filtration in levels, with the level-0 charges represented by Beisert's $\alg{psu}(2|2)$ Lie superalgebra with central extension \cite{Beisert:2005tm,Arutyunov:2006ak}. Almost all level-1 charges have a correspondent level-0 one, except for the secret or bonus symmetry \cite{Matsumoto:2007rh,Beisert:2007ty} (see \cite{deLeeuw:2012jf} for a review with further references). 


The absence of a level-0 hypercharge symmetry of the $S$-matrix prevents the straightforward enlargement to a $\alg{gl}(2|2)$ Yangian. A more sophisticated construction is necessary \cite{B6}, which utilises the $RTT$ relations. What \cite{B6} has shown (see also \cite{B7}) is that Beisert's $S$-matrix naturally produces upon the $RTT$ recipe the most part of the $AdS_5/CFT_4$ Hopf algebra, including the secret symmetry. 

The exploration of lower-dimensional instances of the $AdS/CFT$ integrable system has revealed among other things how much more complex the situation can be, casting new light on the original $AdS_5/CFT_4$ problem as well. A new host of boost-like symmetries has made its appearance primarily in $AdS_3$, and it have been discovered in $AdS_5$ as well.

Integrability in $AdS_3/CFT_2$, both for the $AdS_3 \times S^3 \times S^3 \times S^1$ and the $AdS_3 \times S^3 \times T^4$ background  \cite{Babichenko:2009dk,Sundin:2012gc} (see also \cite{rev3,Borsato:2016hud}), has permitted the adaptation of the largest part of the tools constructed for $AdS_5/CFT_4$, at least in infinite volume, with the Quantum Spectral Curve and full TBA still to be determined. The derivations of the finite-gap equations \cite{OhlssonSax:2011ms} and of the $S$-matrix \cite{Borsato:2012ud,Borsato:2012ss,Borsato:2013qpa,Borsato:2013hoa,Rughoonauth:2012qd,PerLinus,Borsato:2014hja,Borsato:2015mma}, see also \cite{Beccaria:2012kb,Sundin:2013ypa,Bianchi:2013nra,Bianchi:2013nra1,Bianchi:2013nra2}, follow the path of the five-dimensional analysis. However, massless representations make their powerful appearance and force us to dramatically rethink the entire algebraic establishment \cite{Sax:2012jv,Borsato:2016xns,Sax:2014mea,Baggio:2017kza}. The relationship between massless integrable systems and conformal field theory makes its unavoidable entrance in the conversation \cite{Zamol2,Fendley:1993jh,DiegoBogdanAle}, after having realised the inadequacy of the standard massive techniques to deal with the massless sector \cite{Lloyd:2013wza,Abbott:2012dd,Abbott:2014rca,MI,Abbott:2020jaa} (see also \cite{Eberhardt:2017fsi,Gaber1,Gaber2,Gaber3,deLeeuw:2020ahe,Gaber4,Gaber5,
GaberdielUltimo,Prin,Prin1,Abbott:2015mla,Per9,Hoare:2018jim,Pittelli:2014ria,Regelskis:2015xxa,Pittelli:2017spf,JuanMiguelAle}).

The clearest manifestation of the new type of symmetries is obtained by thinking of their existence as a deformation, in the quantum group sense, of the natural Poincar\'e supersymmetry of the original string sigma model, which is lost upon gauge-fixing. In $AdS_5 \times S^5$ this idea started with \cite{CesarRafa,Charles} (see also \cite{Pachol:2015mfa}) and brought to the recent findings of \cite{BorAle}. In the case of massless $AdS_3/CFT_2$ excitations, a series of works established a large number of surprising results \cite{JoakimAle,BorStromTorri,Fontanella:2016opq}. Ultimately, in \cite{FontanellaTorrielli,Fontanella:2019ury} a change of variables was discovered which recasts the non-relativistic $S$-matrix for massless left-left and right-right moving modes, with the inclusion of the dressing factor, in manifestly difference-form. Furthermore, the exact same functional form applies to the non-relativistic as well as to the relativistic $S$-matrix as obtained in \cite{DiegoBogdanAle}. This has allowed to write the massless TBA in perfect analogy to \cite{DiegoBogdanAle} also in the non-relativistic limit \cite{Fontanella:2019ury}. 

Here we will revisit the $\alg{su}(1|1)^2$ algebra\footnote{We shall always understand that our algebras are superalgebras.} that describes the massless sector of the $AdS_3 \times S^3 \times M^4$ scattering problem and the modified Poincar\'e structure which we can introduce to describe the remnant of Poincar\'e symmetry after gauge-fixing the reparameterisation freedom of the string sigma model. In \cite{Creature} we were able to construct a consistent coproduct for the boost operator associated to the modified Poincar\'e algebra that was valid for short representations. In order to understand how to write a coproduct map for the boost operator, we will study the interplay between the two $\alg{su}(1|1)$ subalgebras by making them depend on different momenta. We will study the consistency conditions that this new Poincar\'e structure has to satisfy, which amounts to checking if the generators satisfy the Jacobi identity. This will provide us with six different possible algebras, which we proceed to study in some detail. With that information we will give the form of the coproduct for the boost operator for each of the different algebras.

The article is structured as follows. In Section 2 we review  the $\alg{su}(1|1)^2$ algebra and its outer automorphisms group. We will also study the modified Poincar\'e algebra built upon it, assuming in this case that each copy has a different momentum, and check which restrictions we have to impose in order to get a consistent algebra. Section 3 is devoted to describing each of these consistent algebras separately. In Section 4 we will construct the coproduct of the boost operator associated to the modified Poincar\'e  algebra for each of the different cases we described in Section 3. In Section 5 we summarise our results and present some concluding comments.

\section{\label{AdS3}q-Poincaré algebra in $AdS_3/CFT_2$}

\subsection{The algebra}
In this article we will be concerned with the massless sector of the $AdS_3 \times S^3 \times M_4$ superstring theory, so we will study (one copy of) the $\alg{su}(1|1)^2 \equiv \alg{su}(1|1)_L \oplus \alg{su}(1|1)_R$ scattering problem. However, in contrast with what is usually done, we are going to consider the case where the momentum associated with each of the $\alg{su}(1|1)$ algebras involved is in principle different: $p_L$ not necessarily equal to $p_R$. The generators of the resulting algebra, together with the two independent associated boost generators, satisfy the following (anti-)commutation relations\footnote{The odd vs. even grading of the generators can be evinced by the type of anti-commutation vs. commutation relations they satisfy.}
\begin{gather}
\label{algeq}
\{\alg{Q}_A, \alg{S}_A\} \ = \ \alg{H}_A \ ,  \quad [\alg{J}_A, p_A] \ = \ i \alg{H}_A \ , \quad  [\alg{J}_A, \alg{H}_A] \ = i \alg{H}_A \Phi_A 
\ ,\nonumber\\
[\alg{J}_A, \alg{Q}_A] \ = \phi_{A}^Q \, \alg{Q}_A \ , \qquad [\alg{J}_A, \alg{S}_A] = \phi_{A}^S \, \alg{S}_A \ ,
\end{gather}
where {\footnotesize $A = L,R$}, and we have only reported explicitly the non vanishing (anti-)commutators. 
The generators $\alg{Q}_A$ and $\alg{S}_A$ are of fermionic (odd) nature ({\it supercharges}) and form the odd part of $\alg{su}(1|1)_L \oplus \alg{su}(1|1)_R$, while the generators $\alg{H}_A$ ({\it energies}), $\alg{J}_A$ ({\it boosts}) and $p_A$ ({\it momenta}) are of bosonic (even) nature. Such generators come from extending $\alg{su}(1|1)_L \oplus \alg{su}(1|1)_R$ in order to incorporate a deformed Poincar\'e subalgebra, with the added novelty that we shall now introduce an $L$ and an $R$ copy of every single generator appearing.
By $\phi_A^Q$, $\phi_A^S$ and $\Phi_A$ we have denoted six non-identically-vanishing functions of the momentum generators $p_A$. Their form, as we will amply discuss, is restricted by the Jacobi identities, although not completely fixed: for instance, one has
\begin{equation}
i \alg{H}_A \Phi_A \ = [\alg{J}_A, \alg{H}_A] \ = [\alg{J}_A, \{ \alg{Q}_A , \alg{S}_A \} ] =\{ [\alg{J}_A, \alg{Q}_A ], \alg{S}_A\}+\{ [\alg{J}_A, \alg{S}_A ], \alg{Q}_A\} \ = (\phi_{A}^Q +\phi_{A}^S ) \alg{H}_A \ . \label{phirestriction}
\end{equation}

Apart from the generators we have introduced, we are interested in centrally-extending the algebra in the following way
\begin{equation}
\label{algeqc}
\{\alg{Q}_{\smallL}, \alg{Q}_{\smallR}\} \ = \ \alg{P} \ , \qquad  \{\alg{S}_{\smallL}, \alg{S}_{\smallR}\} \ = \ \alg{K}  \ .
\end{equation}

The algebra so defined is traditionally denoted by $\alg{su}(1|1)^2_{c.e.}$, where $c.e.$ indicates the central extension.

\subsubsection{Outer Automorphisms}\label{subaut}

The centrally-extended $\alg{su}(1|1)^2$ algebra (i.e., without the boost and momentum generators) admits a large outer automorphism group. One particularly prominent generator among the outer automorphisms is the so-called {\it hypercharge}, denoted by $\mathfrak{B}$, which acts on the fermionic generators as
\begin{align*}
[\mathfrak{B},\mathfrak{Q}_L] &= 2 i\mathfrak{Q}_L \ , & [\mathfrak{B},\mathfrak{S}_L] &= - 2i \mathfrak{S}_L \ , & [\mathfrak{B},\mathfrak{Q}_R] &= -2 i\mathfrak{Q}_R \ , & [\mathfrak{B},\mathfrak{S}_R] &= 2i \mathfrak{S}_R \ .
\end{align*}

As described in \cite{Regelskis:2015xxa}, the full set of outer automorphisms form a $GL(2)^2$ group that rotates the fermionic generators with the same quantum number under the hypercharge $\alg{B}$. This group is defined by the action
\begin{align}
	\begin{pmatrix}
	\alg{Q}_L \\
	\alg{S}_R	
	\end{pmatrix} &\longmapsto \lambda \begin{pmatrix}
	\alg{Q}_L \\
	\alg{S}_R	
	\end{pmatrix} \ , & \begin{pmatrix}
	\alg{S}_L \\
	\alg{Q}_R	
	\end{pmatrix} &\longmapsto \rho \begin{pmatrix}
	\alg{S}_L \\
	\alg{Q}_R	
	\end{pmatrix} \ ,
\end{align}
where both $\lambda$ and $\rho$ are $GL(2)$ matrices. The action of the outer automorphism group on the central elements is derived from the action on the supercharges. The algebra action associated to this group is then given by
\begin{align}
	[t^\lambda_0 , \alg{Q}_L] &= [t^\lambda_3 , \alg{Q}_L] = \alg{Q}_L \ , & [t^\lambda_0 , \alg{S}_R] &= -[t^\lambda_3 , \alg{S}_R]= \alg{S}_R \ , & [t^\lambda_+ , \alg{S}_R] &= \alg{Q}_L \ , & [t^\lambda_- , \alg{Q}_L] &= \alg{S}_R \ , \notag \\
	[t^\rho_0 , \alg{Q}_R] &= [t^\rho_3 , \alg{Q}_R] = \alg{Q}_R \ , & [t^\rho_0 , \alg{S}_L] &= -[t^\rho_3 , \alg{S}_L]= \alg{S}_L \ , & [t^\rho_+ , \alg{Q}_R] &= \alg{S}_L \ , & [t^\rho_- , \alg{S}_L] &= \alg{Q}_R \ ,
\end{align}
where $t^\lambda_0$, $t^\lambda_3$ and $t^\lambda_\pm$ and $t^\rho_0$, $t^\rho_3$ and $t^\rho_\pm$ denote the two copies of $\alg{gl}(2)$ Lie algebra generators, while the remaining commutation relations vanish. Notice that the hypercharge can be written in terms of these outer automorphisms as $2i (t^\lambda_0 - t^\rho_0)$.

The action of the outer automorphisms on the boost operators is not closed, which forces us to include (up to six) new boost operators representing the different $[t_\pm^\chi, \alg{J}_A]$ and $[t_\pm^\lambda t_\pm^\rho, \alg{J}_A]$. Because all the properties of these additional generators can be derived from the $\alg{J}_A$ using Jacobi identities, we will only focus on those two out of the eight boosts. In specific representations such boosts are expected to largely collapse onto the two boosts we are studying, modulo perhaps a different momentum-dependent prefactor.

\subsection{The boost and the handedness}

The careful reader might have noticed that we have only enumerated the action of the boost on generators with the same handedness among the non-zero (anti-)commutation relations. An important step we need to take is to generalise this construction to incorporate a non-zero action for the boost of one handedness on generators of the opposite handedness.

Guided by physical input, we will first restrict the {\it energy} generators $\alg{H}_A$ to being non-identically-vanishing positive even functions of the respective momenta, {\it i.e.} $\alg{H}_A = \alg{H}_A(p_A)$. Then, we postulate the action of a boost from one handedness onto a momentum generator from the opposite handedness as
\begin{equation}
	[\alg{J}_L , p_R]=i \alg{H}_L \mathfrak{d}_{LR} \ , \qquad [\alg{J}_R , p_L]=i \alg{H}_R \mathfrak{d}_{RL} \ ,
\end{equation}
where $\mathfrak{d}_{AB}$ are functions of the generators $p_A$. Determining these functions will be one of the main purposes of our work. To do so we will assume that our choice above can be generalised to a relation for all the generators:
\begin{equation}
	\alg{H}_B [\alg{J}_A , X_B] = \alg{H}_A \mathfrak{d}_{AB} [\alg{J}_B , X_B ] \ , \label{keyrelation}
\end{equation}
where $X$ represents any generator with well-defined handedness and $A\neq B$. The action on the two generators that have mixed handedness, i.e. $\alg{P}$ and $\alg{K}$, is inferred from the Jacobi identities involving one boost and two supercharges, e.g. $[\alg{J}_L, \mathfrak{P}] = ( \phi_L^Q \mathfrak{H}_R + \phi_R^Q \mathfrak{H}_L \mathfrak{d}_{LR} ) \mathfrak{P}$.

In order to analyse the consistency of our two-handed algebra, first we have to elaborate on the value of $[\alg{J}_L,\alg{J}_R]$. There are essentially two arguments why only $[\alg{J}_L, \alg{J}_R] = 0$ is a sensible choice in this setting:  Firstly, we want to maintain the underlying $\mathbb{Z}_2$-symmetry given by the $L \leftrightarrow R$ exchange that the algebra has up to this point. This choice will make sure that such symmetry manifests itself throughout our analysis (e.g.  in (\ref{requir})), as a $\mathbb{Z}_2$-symmetry would require $[\alg{J}_L,\alg{J}_R] = [\alg{J}_R,\alg{J}_L]$ for the (bosonic) $\alg{J}_A$.  Secondly, we want (\ref{keyrelation}) to make sense for all algebra elements $X_B$, including the boost operators. Considering (\ref{keyrelation}) with two differently-handed boosts lets us see that $[\alg{J}_L, \alg{J}_R] $ should vanish indeed. We will later revisit this point for the case of a particular representation of the boost.

Now we can study the possible values of $\mathfrak{d}_{AB}$. These two new functions are not arbitrary, but are fixed by the Jacobi identities involving two boost generators of different handedness and a momentum
\begin{equation}
\label{assum}
[ \alg{J}_L , [\alg{J}_R, p_R]] -[\alg{J}_R , [\alg{J}_L , p_R]] + [p_R, [\alg{J}_L , \alg{J}_R]]\ = [ \alg{J}_L , i \alg{H}_R] - [\alg{J}_R , i \alg{H}_L \mathfrak{d}_{LR}] = 0 \ ,
\end{equation}
which can be simplified to
\begin{equation}
	\alg{H}_L \mathfrak{d}_{LR} \Phi_R = -i\alg{H}_L [\alg{J}_R , \mathfrak{d}_{LR}] + \alg{H}_R \mathfrak{d}_{RL} \Phi_L \mathfrak{d}_{LR}\ . \label{eqDelta}
\end{equation}
The other Jacobi identity gives the same equation with the labels $L$ and $R$ exchanged. In addition, we have to consider the Jacobi identities involving the two boost operators and any of the energies, but they give us no extra information. Indeed, we have
\begin{align*}
[\alg{J}_L , [\alg{J}_R , \alg{H}_R]] &= [\alg{J}_R , [\alg{J}_L, \alg{H}_R]] \\
[\alg{J}_L , \alg{H}_R \Phi_R] &= [\alg{J}_R , \alg{H}_L \mathfrak{d}_{LR} \Phi_R] \\
[\alg{J}_L , \alg{H}_R] \Phi_R + \alg{H}_R [\alg{J}_L , \Phi_R] &= [\alg{J}_R , \alg{H}_L \mathfrak{d}_{LR} ] \Phi_R + \alg{H}_L \mathfrak{d}_{LR} [\alg{J}_R , \Phi_R] \\
\alg{H}_L \mathfrak{d}_{LR} \Phi_R^2 &= -i\alg{H}_L [\alg{J}_R , \mathfrak{d}_{LR}] \Phi_R + \alg{H}_R \mathfrak{d}_{RL} \Phi_L \Phi_R \mathfrak{d}_{LR}
\end{align*}
which do not add extra information on top of (\ref{eqDelta}). Jacobi identities involving two boosts and a supercharge are also congruent with (\ref{eqDelta}).

By direct examination of (\ref{eqDelta}) we can already find five different solutions. The first solution is the trivial case $\mathfrak{d}_{RL}=\mathfrak{d}_{LR}=0$. The second solution is $\mathfrak{d}_{RL}=0$ and $\mathfrak{d}_{LR}=\zeta \alg{H}_R$, and the third solution is obtained from the second one by swapping handedness. $\zeta$ is in principle a function of the momentum generators, however it has to be central with respect to the entire algebra, including the boost generators, so the only option is for it to be a constant function independent of $p_A$. The remaining two solutions are given by $\mathfrak{d}_{RL}=\mathfrak{d}_{LR}=\pm 1$.\footnote{As we will more explicitly see later, $\mathfrak{d}_{AB} = \pm 1$ is only an admissible solution for $\alg{H}_L \propto \alg{H}_R$.}

We now prove that we can only have two categories of algebras, namely either i) $\mathfrak{d}_{LR} \mathfrak{d}_{RL}=1$ or ii) one (or both) $\mathfrak{d}_{AB}=0$. We can use the above equations and equation~(\ref{keyrelation}) to compute the boosts' action on $\mathfrak{d}_{LR} \mathfrak{d}_{RL}$, from which we get 
\begin{align}
	-i\alg{H}_L [\alg{J}_R , \mathfrak{d}_{LR} \mathfrak{d}_{RL} ] =\left( \alg{H}_L \Phi_R - \alg{H}_R \Phi_L \mathfrak{d}_{RL} \right) \mathfrak{d}_{LR} \mathfrak{d}_{RL} (1-\mathfrak{d}_{LR} \mathfrak{d}_{RL} ) \ , \notag \\
	-i\alg{H}_R [\alg{J}_L , \mathfrak{d}_{LR} \mathfrak{d}_{RL} ] =\left( \alg{H}_R \Phi_L - \alg{H}_L \Phi_R \mathfrak{d}_{LR} \right) \mathfrak{d}_{LR} \mathfrak{d}_{RL} (1-\mathfrak{d}_{LR} \mathfrak{d}_{RL} ) \ . \label{DeltaDeltaevolution}
\end{align}
Imposing the further consistency $\alg{H}_L [\alg{J}_R , \mathfrak{d}_{LR} \mathfrak{d}_{RL} ]=\mathfrak{d}_{RL} \alg{H}_R [\alg{J}_L , \mathfrak{d}_{LR} \mathfrak{d}_{RL} ]$ we get that
\begin{equation}
\label{inhere}
	\left[ \alg{H}_L \Phi_R (1-\mathfrak{d}_{LR} \mathfrak{d}_{RL}) - \alg{H}_R \Phi_L ( \mathfrak{d}_{RL} - \mathfrak{d}_{LR}) \right] \mathfrak{d}_{LR} \mathfrak{d}_{RL} (1-\mathfrak{d}_{LR} \mathfrak{d}_{RL})= 0 \ ,
\end{equation}
and a similar condition for $\alg{H}_R [\alg{J}_L , \mathfrak{d}_{LR} \mathfrak{d}_{RL} ]$, which can be obtained by swapping the $L$ and $R$ labels in (\ref{inhere}). These two consistency conditions can be solved simultaneously by either $\mathfrak{d}_{LR}=0$, $\mathfrak{d}_{RL}=0$ or $\mathfrak{d}_{LR} \mathfrak{d}_{RL}=1$. One can show that no further solutions come from requiring 
\begin{eqnarray}
\label{requir}
&&\alg{H}_L \Phi_R (1-\mathfrak{d}_{LR} \mathfrak{d}_{RL}) - \alg{H}_R \Phi_L ( \mathfrak{d}_{RL} - \mathfrak{d}_{LR})=0 \qquad \nonumber\\
\mbox{and}  &&\alg{H}_R \Phi_L (1-\mathfrak{d}_{LR} \mathfrak{d}_{RL}) - \alg{H}_L \Phi_R ( \mathfrak{d}_{LR} - \mathfrak{d}_{RL})=0
\end{eqnarray}
simultaneously, which would descend from assuming that $\mathfrak{d}_{LR} \mathfrak{d}_{RL} (1-\mathfrak{d}_{LR} \mathfrak{d}_{RL})$ is non-zero. In fact, after some manipulations, we can see that the only values that solve both equations (\ref{requir}) simultaneously are $\mathfrak{d}_{LR}=\mathfrak{d}_{RL}=\pm 1$.

Category ii) reproduces the first three solutions described below (\ref{eqDelta}), while category i) includes further solutions in addition to the fourth and fifth described there. Specifically, within category i) we can reduce the constraint on $\mathfrak{d}_{LR}$ to
\begin{equation}
	-i\alg{H}_L [\alg{J}_R , \mathfrak{d}_{LR}]= \alg{H}_L \mathfrak{d}_{LR} \Phi_R - \alg{H}_R \Phi_L \ .
\end{equation}
A first step to solve this equation is to take out a factor of $\alg{H}_R$ from $\mathfrak{d}_{LR}$. If we define $\mathfrak{d}_{LR}=\bar{\mathfrak{d}}_{LR} \alg{H}_R$, with $\bar{\mathfrak{d}}_{LR}$ a function of the momenta, we can see that
\begin{align*}
	 -i\alg{H}_L [\alg{J}_R , \mathfrak{d}_{LR}]=-i\alg{H}_L \alg{H}_R [\alg{J}_R , \bar{\mathfrak{d}}_{LR}]+ \alg{H}_L \alg{H}_R \bar{\mathfrak{d}}_{LR} \Phi_R = \alg{H}_L \alg{H}_R \bar{\mathfrak{d}}_{LR} \Phi_R - \alg{H}_R \Phi_L \ .
\end{align*}
Thus $i\alg{H}_L [\alg{J}_R , \bar{\mathfrak{d}}_{LR}]= \Phi_L$, which notably simplifies the constraint. Finally, let us prove that $\bar{\mathfrak{d}}_{LR} \alg{H}_L$ should commute with $\alg{J}_L$, fixing it to be a central element with respect to all the elements of the algebra, including the boost:
\begin{align*}
	\alg{H}_R [\alg{J}_L, \bar{\mathfrak{d}}_{LR} \alg{H}_L] & =\alg{H}_R [\alg{J}_L, \bar{\mathfrak{d}}_{LR}] \alg{H}_L+ \alg{H}_R \bar{\mathfrak{d}}_{LR} [\alg{J}_L, \alg{H}_L]= \bar{\mathfrak{d}}_{LR} \alg{H}_L [\alg{J}_R, \bar{\mathfrak{d}}_{LR}]  \alg{H}_L+ i \alg{H}_R \bar{\mathfrak{d}}_{LR} \alg{H}_L \Phi_L \\
	& = -i \bar{\mathfrak{d}}_{LR} \alg{H}_R \Phi_L \alg{H}_L+ i \alg{H}_R \bar{\mathfrak{d}}_{LR} \alg{H}_L \Phi_L=0 \ .
\end{align*}
We can prove in a similar way that $\bar{\mathfrak{d}}_{LR} \alg{H}_L$ commutes with $\alg{J}_R$. Thus, we finally obtain the relation
\begin{equation}
	\alg{H}_L \mathfrak{d}_{LR} = \zeta \alg{H}_R \ ,
\end{equation}
where $\zeta$ is a constant. This establishes the sixth type of algebras we shall study. As we will explain in Section \ref{S3}, there is a nomenclature that lends itself to our situation nicely, namely calling the algebras either \textit{separable} or \textit{differential}:\footnote{During the latter part of our discussion, we will indeed find that the case of $\alg{H}_L \mathfrak{d}_{LR} = \zeta \alg{H}_R$ has connections to both the differential and separable algebras.}
\begin{equation*}
  \text{\textit{separable} algebras} =
    \begin{cases}
      &\mathfrak{d}_{AB} = 0\\
      &\mathfrak{d}_{LR} = 0 \textit{ and } \mathfrak{d}_{RL} = \zeta \alg{H}_L\\
      &\mathfrak{d}_{RL} = 0 \textit{ and } \mathfrak{d}_{LR} = \zeta \alg{H}_R
    \end{cases}       \qquad
  \text{\textit{differential} algebras} =
    \begin{cases}
     &\mathfrak{d}_{AB} = +1\\
     &\mathfrak{d}_{AB} = -1\\
     & \alg{H}_L \mathfrak{d}_{LR} = \zeta \alg{H}_R
    \end{cases} \ .
\end{equation*}

\subsection{The differential representation}

The most natural representation from a physical standpoint is the one where we realise the momenta as real variables, and the boost operator as a derivative operator:
\begin{equation}
	\alg{J}_A= i \alg{H}_A \, \frac{d\hphantom{ p_A}}{d p_A} \ . \label{differentialformboost}
\end{equation}
The derivative is understood in a convective fashion, for example
\begin{eqnarray}
\label{rewr}
\frac{d \hphantom{ p_A}}{d p_R}=\frac{\partial \hphantom{ p_A}}{\partial p_R} + \frac{d p_L}{d p_R} \frac{\partial \hphantom{ p_A}}{\partial p_L} \ .
\end{eqnarray} 
This is especially relevant for $\mathfrak{d}_{AB}$, $\alg{P}$ and $\alg{K}$, as they depend explicitly on both momenta. In this representation we can understand $\Phi_A=\alg{H}'_A=\frac{d \alg{H}_A}{d p_A}$, and $\mathfrak{d}_{LR}$ and $\mathfrak{d}_{RL}$ are nothing but the derivative of one momentum with respect to the other, where we understand that eventually one can choose either of the $p_A$ to be the only independent variable. In this respect, $\mathfrak{d}_{AB}$ assume the rôle of Jacobians. This allows us to rewrite (\ref{rewr}) as
\begin{eqnarray}
\frac{d \hphantom{ p_A}}{d p_R}=\frac{\partial \hphantom{ p_A}}{\partial p_R} + \mathfrak{d}_{RL} \frac{\partial \hphantom{ p_A}}{\partial p_L} \ .
\end{eqnarray}

In this representation we can also write equation (\ref{eqDelta}) and its $L\leftrightarrow R$ symmetric as
\begin{equation}
	\alg{H}_L\alg{H}_R\frac{d \mathfrak{d}_{LR}}{d p_R}= \left( \alg{H}_L \Phi_R -\alg{H}_R\Phi_L \mathfrak{d}_{RL} \right) \mathfrak{d}_{LR} \ , \qquad \alg{H}_L\alg{H}_R\frac{d \mathfrak{d}_{RL}}{d p_L}=\left( \alg{H}_R \Phi_L -\alg{H}_L\Phi_R \mathfrak{d}_{LR} \right) \mathfrak{d}_{RL} \ . \label{diffeqDelta}
\end{equation}
The six solutions naturally acquire an induced representation. Given our definition of $\mathfrak{d}_{AB}$, we can see that the first solution presented in the previous section, namely $\mathfrak{d}_{LR}=0=\mathfrak{d}_{RL}$, is the case where $p_R$ and $p_L$ are independent of each other. The second and third solutions imply instead $p_R =  \pm p_L$ and $\alg{H}_R (p_R=\pm p_L) =$ const. $ \alg{H}_L$.

In addition, this differential representation can be used to construct the action of the outer automorphisms on the boost operator, as it reduces to computing the action of the outer automorphisms on $\alg{H}_A$, as e.g.
\begin{align*}
[t_-^\rho, \alg{J}_L] = [t_-^\rho, i \alg{H}_L \frac{d}{d p_L}] = i [t_-^\rho, \alg{H}_L] \frac{d}{d p_L} ,
\end{align*}
where we have $[t_-^\rho, \alg{H}_L] = \alg{P}$.

As a sanity check of the statement made in the previous section about the vanishing of $[\alg{J}_L, \alg{J}_R]$, we quickly want to demonstrate that this is fulfilled within the differential representation framework. The differential representation for the boost features a convective derivative as in (\ref{rewr}), and as $\mathfrak{d}_{AB}$ can depend on the momenta $p_A, p_B$, the expression for $[\alg{J}_L, \alg{J}_R]$ also features such derivatives, as for example
\begin{align*}
[\alg{J}_L, \alg{J}_R] \Big|_{\partial_L \text{  coeff.}} = \alg{H}_L \alg{H}_R (\partial_{p_L} \mathfrak{d}_{RL}) +\alg{H}_L \mathfrak{d}_{LR} (\partial_{p_R} \alg{H}_R) \mathfrak{d}_{RL} + \alg{H}_L \mathfrak{d}_{LR} \alg{H}_R (\partial_{p_R} \mathfrak{d}_{RL}) -\alg{H}_R \mathfrak{d}_{RL} (\partial_{p_L} \alg{H}_L) ,
\end{align*}
which is exactly (the right) equation (\ref{diffeqDelta}) after expressing the convective derivative in terms of partial derivatives - and thus vanishes. The vanishing of the $\partial_{p_R}$ coefficient can be proven in a similar fashion, as it has the same structure but with the $L \leftrightarrow R$ handedness swapped.\par
Above, we were able to see that a decisive difference of the convective differential with respect to the ordinary (holonomic) partial derivatives is that two convective derivatives might not commute due to the possible momentum-dependence of $\mathfrak{d}_{AB}$. However, this in turn is crucial for the vanishing of $[\alg{J}_L, \alg{J}_R] = 0$.

\section{The different acceptable algebras}
\label{S3}

In this section we will substitute the acceptable values of the operators $\mathfrak{d}_{AB}$ we found above and study the different algebras we obtain in the process, which we have divided into two categories. On the one hand, we denote the cases where at least one $\mathfrak{d}_{AB}=0$ as \emph{separable algebras}, as in those cases we can completely decompose them into two independent subalgebras. On the other hand, we denote the other two cases as \emph{differential algebras} because the relation $\mathfrak{d}_{AB} \mathfrak{d}_{BA}=1$ can be understood as the inverse function theorem and $\mathfrak{d}_{AB}$ can be interpreted as the Jacobian of a change of variables from $p_B$ to $p_A$, building upon the intuition provided by the differential representation presented in the previous section.

\subsection{Separable algebras}

Let us first examine the case where we set $\mathfrak{d}_{LR}=\mathfrak{d}_{RL}=0$. The commutation relations involving the boost operators take the form
\begin{align}
[\alg{J}_A, p_A] &= i \alg{H}_A \ , & [\alg{J}_A, \alg{H}_A] &= i \alg{H}_A \Phi_A \ , \notag \\
[\alg{J}_{L}, p_L] &= \alg{H}_L \ , & [\alg{J}_{R}, p_R] &= \alg{H}_R \ , \notag \\
[\alg{J}_A, \alg{Q}_A] & = \phi_{A}^Q \, \alg{Q}_A \ , & [\alg{J}_A, \alg{S}_A] &= \phi_{A}^S \, \alg{S}_A \ , \notag \\
[\alg{J}_{L}, X_R] &= 0 \ , & [\alg{J}_{R}, X_L] &= 0 \ , \notag \\
[\alg{J}_{A}, \alg{P}]&= \phi_A^Q \alg{P} \ , & [\alg{J}_{A}, \alg{K}]&= \phi_A^S \alg{K} \ , \label{separatedalgebra}
\end{align}
where $X_A$ represents any generator with well-defined handedness $A$. We can understand these algebras, by looking at their differential representation, as the ones in which all generators with left handedness only depend on $p_L$ for all representations, and similarly for the right handedness, as the two momenta have to be independent.

We want to better illustrate this separation of the algebra into two by looking at the particular case of the relativistic dispersion relation $\alg{H}_A^2=p_A^2 + m^2$. If we focus on the subalgebra formed by the boosts, the momenta and the energies, we can see that it reduces to a known finite-dimensional Lie algebra because of the constraint $\alg{H}_A \Phi_A=p_A$. In that case we can compare it with the classification of solvable 6-dimensional Lie algebras presented in \cite{Turk}. For that we pick the basis
\begin{align*}
x_{1,2} &= - i \mathfrak{J}_{L,R} \\
n_{1,2} &= \mathfrak{H}_{R,L} - p_{R,L} \\
n_{3,4} &= \mathfrak{H}_{R,L} + p_{R,L},
\end{align*}
where the $n_i$ span the nilradical (which in this case forms an abelian subalgebra) and the $x_i$ form its complement. With this, it is evident that our algebra corresponds to $N_{6,1}^{\alpha \beta \gamma \delta}$ with $\alpha = \delta = 0$ and $\beta = \gamma = -1$ in the notation of said paper. However, one caveat is that an algebra of type $N_{6,1}^{\alpha \beta \gamma \delta}$ is indecomposable provided $\gamma^2 + \delta^2 \neq 0$ and $\alpha \beta \neq 0$, the latter of which is not the case for us. Thus, our algebra actually corresponds to a simple direct sum of the 3-dimensional left-handed and right-handed sides. This is also the situation for the choice $\alg{H}_A \Phi_A \propto [p_A]_q$, which corresponds to the magnonic dispersion relation for a particular value of $q$, i.e., $\alg{H}^2_A = h_A^2 \sin^2  \frac{p_A}{2} + m^2$, with $h_A$ a constant.

We also want to address the question of fixing the functions $\phi_{A}^Q $ and $\phi_{A}^S$ in this setting. On the one hand, the Jacobi identity involving one boost operator and two supercharges with different handedness imposes the relations in the fifth line of~(\ref{separatedalgebra}) thanks to the decoupling of the left and right sectors. This restriction can be combined with equation~(\ref{phirestriction}) to further constraint the form of the central elements. In particular
\begin{equation}
	i \alg{K} \alg{P} \Phi_A= \alg{K} \left( \phi_{A}^Q \alg{P} \right) +\alg{P} \left( \vphantom{\phi_{A}^Q} \phi_{A}^S \alg{K} \right) =\alg{K} [\alg{J}_A , \alg{P} ] + \alg{P} [\alg{J}_A , \alg{K} ]\ ,
\end{equation}
as $\Phi_A$ only depends on $p_A$ while $\alg{P}$ and $\alg{K}$ depend on both momenta. On the other hand, the Jacobi identity involving two different boost and a supercharge imposes the restriction
\begin{equation}
\label{restricc}
	[\alg{J}_A , \phi^Q_B]=[\alg{J}_A , \phi^S_B]=0 \ ,
\end{equation}
for $A \neq B$. In the differential representation this can be interpreted as restricting the functions $\phi^S_A$ and $\phi^Q_A$ to depend only on the momentum $p_A$. This is truly a very heavy restriction, as it forces the central elements to be separable, i.e., $\alg{P}=\alg{P}_R \alg{P}_L$ and $\alg{K}=\alg{K}_R \alg{K}_L$, for this algebra to be consistent. In the differential representation this is immediate, as
\begin{align*}
i \mathfrak{H}_A \frac{d}{d p_A} \alg{P} = [ \alg{J}_A, \mathfrak{P}] = \phi_A^Q \mathfrak{P} \implies i \frac{d}{d p_A} \log \mathfrak{P} = \mathfrak{H}_A^{-1} \phi_A^Q,
\end{align*}
where the latter term \emph{only} depends on $p_A$. Even if $\alg{J}_A$ is a priori not of differential form, the same argument can be made as (\ref{restricc}) still holds since the adjoint action of $\alg{J}_A$ only vanishes on $p_A$-independent elements because of $\mathfrak{d}_{AB}=0$. Thus, since $\alg{J}_A$ vanishes on $\mathfrak{P}^{-1} [\alg{J}_B, \mathfrak{P}]$ for $A \neq B$, the argument follows.

Let us point out that the $\mathfrak{d}_{LR}=0=\mathfrak{d}_{RL}$ case has no applications to $AdS_3$ physics. If we wanted to study massless excitations in $AdS_3$, we would necessarily have to take the moves from central elements of the form $\alg{K} \propto \alg{P} \propto \sin (\frac{p_L + p_R}{4} )$, which do not fulfil the condition of separability.

Let us move our attention to the case $\mathfrak{d}_{LR}=0$ and $\mathfrak{d}_{RL}=\zeta \alg{H}_L$. This case can be transformed back to the $\mathfrak{d}_{LR}=\mathfrak{d}_{RL}=0$ case - after a redefinition of one of the boosts in terms of the $\mathfrak{d}_{AB}=0$ boosts $\alg{J}^0_A$
\begin{equation}
\label{arr1}
 \alg{J}_R (\mathfrak{d}_{RL}=\zeta \alg{H}_L )=\alg{J}_R (\mathfrak{d}_{AB} = 0)-\zeta \alg{H}_R \alg{J}_L (\mathfrak{d}_{AB} = 0) \ ,
\end{equation}
so all of the above arguments apply here with minimal changes. Similarly can be done for the one obtained by swapping the handedness, $\mathfrak{d}_{RL}=0$ and $\mathfrak{d}_{LR}=\zeta \alg{H}_R$, with the redefinition
\begin{equation}
\label{arr2}
 \alg{J}_L (\mathfrak{d}_{LR}=\zeta \alg{H}_R)=\alg{J}_L (\mathfrak{d}_{AB} = 0)-\zeta \alg{H}_L \alg{J}_R (\mathfrak{d}_{AB} = 0) \ .
\end{equation}
For the $\mathfrak{d}_{AB} = 0$ case it also follows that $[\alg{J}_L,\alg{J}_R] = 0$, by virtue of the relationship amongst $L$-handed and $R$-handed generators in this case. (\ref{arr1}) and (\ref{arr2}) imply the same conclusion for $\mathfrak{d}_{LR}=0$ and $\mathfrak{d}_{RL}=\zeta \alg{H}_L$ as well as $\mathfrak{d}_{RL}=0$ and $\mathfrak{d}_{LR}=\zeta \alg{H}_R$, respectively.

\subsection{The $\mathfrak{d}_{LR}=\mathfrak{d}_{RL}=\pm 1$ algebra}

As we have pointed out earlier by referring to the differential representation, this choice of $\mathfrak{d}_{AB}$ imposes that either $p_L=p_R+q$ or $p_L = -p_R+q$ for some constant $q \in \mathbb{R}$, which means that either the two copies built on each $\alg{su} (1|1)$ are exactly the same or one is the parity-transformed of the other. The constant $q$ can be interpreted as our freedom to parameterise the origin of the two momenta in different ways. Thus, we will set it to zero from now on. We should recall that this solution only exists provided
\begin{eqnarray}
\label{ide}
\alg{H}_R \left[ p_R (p_L) \right]=\mathfrak{d}_{LR} \frac{h_R}{h_L} \alg{H}_L (p_L),
\end{eqnarray}
with $h_A$  constants such that $\text{sign} (\mathfrak{d}_{AB} \frac{h_B}{h_A}) = +1$. This is a consequence of the chain of arguments related to (\ref{eqDelta}), leading also to the specialisation (\ref{diffeqDelta}) and the discussion below it. Although this case is very restrictive, the algebra we will present in the next section is a generalisation of this algebra without such constraints. Representations of this particular algebra have been profusely studied for the case $p_L=p_R$, for example in \cite{Regelskis:2015xxa}.

After performing the identifications (\ref{ide}), the algebra can be written as
\begin{gather}
\label{algeqD=1}
\{\alg{Q}_A, \alg{S}_A\} \ = \ \alg{H}_A \ ,  \quad [\alg{J}_A, p_A] \ = \ i \alg{H}_A \ , \quad  [\alg{J}_A, \alg{H}_A] \ = i \alg{H}_A \Phi_A 
\ ,\nonumber\\
[\alg{J}_A, \alg{Q}_A] \ = \phi_{A}^Q \, \alg{Q}_A \ , \qquad [\alg{J}_A, \alg{S}_A] = \phi_{A}^S \, \alg{S}_A \ , \nonumber \\
\{\alg{Q}_{\smallL}, \alg{Q}_{\smallR}\} \ = \ \alg{P} \ , \qquad  \{\alg{S}_{\smallL}, \alg{S}_{\smallR}\} \ = \ \alg{K} \ , \nonumber \\
h_A [\alg{J}_B , \alg{P}] \ = \left( h_L \phi_{R}^Q + h_R \phi_{L}^Q \right) \alg{P} \ , \qquad h_A [\alg{J}_B , \alg{K}] \ = \left( h_L \phi_{R}^S + h_R \phi_{L}^S \right) \alg{K} \ , \notag \\
[\alg{J}_A , X_B]= \frac{h_A}{h_B} [\alg{J}_B, X_B] \ ,
\end{gather}
where {\footnotesize $A,B = L,R$}, {\footnotesize $A\neq B$} and $X$ is any generator with well-defined handedness. Notice that we have made explicit use of the condition $h_A \alg{H}_B = \mathfrak{d}_{AB} h_B \alg{H}_A$. The above relations also imply $[\alg{J}_L, \alg{J}_R] = 0$, as $\alg{J}_L$ and $\alg{J}_R$ essentially coincide for $\mathfrak{d}_{AB} = \pm 1$.

Let us consider a special case of this algebra. We first focus again on the 6-dimensional algebra formed by the (Poincaré-like) generators $\mathfrak{J}_A$, $\mathfrak{H}_A \text{ and } p_A$ with relativistic dispersion relation $\alg{H}_A^2 = p_A^2 + m^2$. Imposing the relation between our energies, we can reduce this 6 dimensional algebra to a 5-dimensional one by the identification $\alg{H}_R= \alg{H}_L=\alg{H}$, formally spanned by $\{ \alg{J}_L, \alg{J}_R, p_L, p_R, \alg{H} \}$. Imposing now the constraint $\mathfrak{d}_{AB} = \pm 1$, we find this algebra to have a 2-dimensional centre spanned by the generators $p_L \mp p_R$ and $\alg{J}_L - \alg{J}_R$, respectively for either of the cases. Notice that $p_L \mp p_R$ parametrises the additional shift between the momenta that we commented above. We can canonically project and mod this center out, leaving us with the 3-dimensional algebras spanned by $\{ \alg{J}_L + \alg{J}_R, p_L \pm p_R, \alg{H} \}$, respectively. In the case $\mathfrak{d}_{LR} = \mathfrak{d}_{RL} = 1$, the remaining algebra obtained after we mod-out the centre has the following non-vanishing commutation relations
\begin{align*}
	\left[ \frac{\alg{J}_L + \alg{J}_R}{2} , \frac{p_L + p_R}{2} \pm \alg{H} \right]=\pm \left( \frac{p_L + p_R}{2} \pm \alg{H} \right) \ ,
\end{align*}
which correspond to one of the four irreducible 3-dimensional solvable algebras \cite{graaf}. This construction can be extended to different dispersion relations, thanks to the restriction $\alg{H}_R=\alg{H}_L$ which keeps $p_L \mp p_R$ and $\alg{J}_L - \alg{J}_R$ as central elements. 

We shall now return to study general features of the $\mathfrak{d}_{LR} = \mathfrak{d}_{RL} =\pm 1$ algebra.

\subsubsection{Differential Representation\label{repo}}

Let us consider the differential representation for the  $\mathfrak{d}_{LR} = \mathfrak{d}_{RL} =\pm 1$ algebra. Furthermore, with an eye towards applications to massless excitations in $AdS_3$, we are going to choose the following particular dependence on the momenta for the central generators
\begin{equation}
\label{fit}
	\alg{H}_L=h_L \, \bigg\vert \sin \frac{p_L}{2}\bigg\vert \, \alg{1} \ , \qquad \alg{H}_R= \mathfrak{d}_{LR} \, h_R \, \bigg\vert\sin \frac{p_R}{2} \bigg\vert \, \alg{1} \ , \qquad \alg{K} \propto \alg{P} \propto \bigg \vert\sin \left( \frac{p_L + \mathfrak{d}_{LR} p_R}{4} \right) \bigg\vert \, \alg{1}  \ ,
\end{equation}
where $\alg{1}$ is the identity matrix in $(1 \mbox{boson},1 \mbox{fermion})$-dimensional space, and $h_L$ and $h_R$ are positive real numbers. The quantity $h_L + h_R$ is usually identified as a coupling constant appearing in the dispersion relation of the fundamental excitations in the context of the $AdS_3/CFT_2$ duality. Moreover, we set the two boost to be equal up to a multiplicative constant $h_R \alg{J}_L\equiv h_L \alg{J}_R$, which is consistent with the constraints imposed by the Jacobi identities specialised to the algebra in object. These two identifications together with the constraint $p_L=\pm p_R$ also imply $\mathfrak{d}_{LR} h_R \alg{H}_L = h_L \alg{H}_R$ in this representation.


We now restrict ourselves to
\begin{eqnarray}
\label{same}
p_L \equiv  p_R \equiv p >0
\end{eqnarray} 
in order not to have to deal with the absolute value in the expression for the energies. Physically, this amount to restricting to the so-called "right-moving" representation. Regarding the supercharges, we find that it is possible to set $\phi^Q=\phi^S$, so that in this representation we have
\begin{align}
[\alg{J}_A, \alg{Q}_B] \ = \frac{i \Phi_A}{2} \, \alg{Q}_B \ , \qquad [\alg{J}_A, \alg{S}_B] = \frac{i \Phi_A}{2} \, \alg{S}_B \ ,
\end{align}
for any combination of $A$ and $B$. Using such commutation relations, we can strip the supercharges of the dependence on the momentum and write (keeping in mind equation (\ref{same}))
\begin{align}
\alg{Q}_L &= \sqrt{\alpha h_L \sin \frac{p}{2}} \hat{Q}_L \ , & \alg{Q}_R &= \sqrt{\beta h_R \sin \frac{p}{2}}  \hat{Q}_R \ , \notag \\
\alg{S}_L &=\sqrt{ \frac{h_L}{\alpha} \sin \frac{p}{2}}  \hat{S}_L \ , & \alg{S}_R &=  \sqrt{\frac{h_R}{\beta} \sin \frac{p}{2}}  \hat{S}_R \ , \label{hatsupercharges}
\end{align} 
where $\alpha$ and $\beta$ are constants and the hatted quantities are matrices satisfying
\begin{equation}
 \{ \hat{Q}_L , \hat{S}_L \} = \{ \hat{Q}_R , \hat{S}_R \}=\alg{1} \ ,
\end{equation}
in order for this algebra to agree with our definition of $\alg{H}_A$.

Furthermore, we can repeat the same rewriting with the other two central elements $\alg{P}$ and $\alg{K}$. It is easy to see that
\begin{equation}
	\alg{P}= \bigg\vert h_L h_R\alpha \beta \bigg\vert \sin \frac{p}{2} \hat{P} \ , \qquad \alg{K}= \bigg\vert h_L h_R\frac{1}{\alpha \beta}  \bigg\vert \sin \frac{p}{2} \hat{K} \ .
\end{equation}

Furthermore, centrality imposes that $\hat{P}$ and $\hat{K}$ should be proportional to the identity. Combining that with their definition in terms of supercharges we can write
\begin{equation}
	\{ \hat{Q}_L , \hat{Q}_R \}=\hat{P}=\gamma \alg{1} \ , \qquad \{ \hat{S}_L , \hat{S}_R \}=\hat{K}=\eta \alg{1}
\end{equation}
with $\gamma$ and $\eta$ being constants. The first two equalities come from our definition of $\alg{H}_A$. Although we can perform the redefinitions $\hat{Q}_A\rightarrow \frac{1}{\sqrt{\gamma}} \hat{Q}_A$ and $\hat{S}_A\rightarrow \sqrt{\gamma}\hat{S}_A$, which allows us to set $\gamma=1$ from now on, there is no method to set $\eta$ to $1$ at the same time. In fact, this parameter $\eta$ is actually related to the shortening condition, as short representations can only exist if $\eta=1$. For general values of $\eta$ we can see that the hatted quantities fulfil
\begin{align}
	&\{ (1+x \eta ) \hat{Q}_L - (1+x) \hat{S}_R , x \hat{S}_L + \hat{Q}_R \} = 0 \ , \\
	&\{ y \hat{Q}_L + \hat{S}_R , (y+1) \hat{S}_L - (y+\eta) \hat{Q}_R \} = 0 \ ,
\end{align}
for any values of $x$ and $y$.

Thus, we can write a set of relations which are peculiar to this representation: \footnote{We could also state the algebra without reference to handedness due to the above identification.}
\begin{gather}
\left[ \alg{J}_L + \alg{J}_R , \frac{p_L + p_R}{2} \right]=  \alg{H}_L+\alg{H}_R \ , \qquad \left[ \alg{J}_L + \alg{J}_R , \alg{H}_L+\alg{H}_R \right] \ = i (\alg{H}_L+\alg{H}_R ) (\Phi_L+\Phi_R ) \ , \notag \\
	\{ \hat{Q}_L , \hat{S}_L \} = \{ \hat{Q}_R , \hat{S}_R \}=\{ \hat{Q}_L , \hat{Q}_R \}=\eta^{-1} \{ \hat{S}_L , \hat{S}_R \}=\alg{1} \ ,
\end{gather}
which represents a complete separation of the $\alg{su} (1|1)$ algebra from the modified Poincaré algebra.

\subsection{The $\alg{H}_L \mathfrak{d}_{LR}= \zeta \alg{H}_R$ algebra}

Substituting the value of $\mathfrak{d}_{AB}$ into equation~(\ref{keyrelation}) gives us that the relation between the two handedness of this algebra is
\begin{equation}
\label{twoH}
	[\alg{J}_A , X_B ]=\Big( \zeta + (1-\zeta) \mathfrak{d}_{AB} \Big) [\alg{J}_B , X_B] \ ,
\end{equation}
for any generator $X$ and for any combination of {\footnotesize $A,B = L,R$}. In contrast with the previous case, no linear combination of $p_L$ and $p_R$ is central. Nevertheless, one can readily check that we can indeed construct the boost operators $\alg{J}_L$ and $\alg{J}_R$ of this algebra from the ones of the $\mathfrak{d}_{LR} = \mathfrak{d}_{RL} = 0$ algebra by making the following identifications:
\begin{align}
\alg{J}_L { (\alg{H}_L \mathfrak{d}_{LR}= \zeta \alg{H}_R)}&= {\alg{J}_L (\mathfrak{d}_{AB} = 0)} + \zeta {\alg{J}_R (\mathfrak{d}_{AB} = 0)} \ , \notag  \\
\alg{J}_R { (\alg{H}_L \mathfrak{d}_{LR}= \zeta \alg{H}_R)}&= {\alg{J}_R (\mathfrak{d}_{AB} = 0)} + \zeta {\alg{J}_L (\mathfrak{d}_{AB} = 0)} \  ,\label{reex}
\end{align}
Notice that (\ref{reex}) also implies $[\alg{J}_L, \alg{J}_R] = 0$, as we can just re-express it in terms of commutators of the form $[{\alg{J}_L (\mathfrak{d}_{AB} = 0)}, {\alg{J}_R (\mathfrak{d}_{AB} = 0)}]$, which we know to vanish.

\subsubsection{Differential Representation}

As we are interested in representations defined by $\alg{H}_L (p_L) = h_L \sin \frac{p_L}{2} \alg{1}$ and $\alg{H}_R (p_R) = h_R \sin \frac{p_R}{2} \alg{1}$, substituting these expressions into the expression for $\mathfrak{d}_{LR}$ we get
\begin{equation}
	\mathfrak{d}_{LR}=\partial_{p_L} p_R = \frac{h_R}{h_L} \zeta \csc \frac{p_L}{2} \sin \frac{p_R}{2} \Longrightarrow p_R(p_L)=4\arccot \left( \kappa \cot^{\gamma} \frac{p_L}{4} \right) \ ,
\end{equation}
where $\kappa$ is an integration constant and $\gamma=\frac{h_R}{h_L} \zeta$. Depending on whether $|\kappa|$ is greater or smaller than 1, we have to reduce the range of $p_R$ or the range of $p_L$. This representation amounts to having
\begin{equation}
	\alg{H}_R (p_R (p_L))=\frac{\kappa h_R}{h_L^{\gamma}} \frac{\alg{H}_L^{\gamma}}{\kappa^2 \cos^{2 \gamma} \frac{p_L}{4} +\sin^{2\gamma} \frac{p_L}{4}} \ .
\end{equation}
when $\alg{H}_R$ is written in terms of $p_L$.

Notice that the representation from section \ref{repo} is included in this family of representations as the case $\kappa=\gamma=1$.

\section{Coproduct map for the boost operator for the $\mathfrak{d}_{AB} = \pm 1$ case}\label{universalboost}

In this section we will continue the line of research we started in \cite{Creature}, where we picked a particular representation and coproduct of the $\alg{su}(1|1)^2_{c.e.}$ algebra and constructed a boost operator whose coproduct quasi-cocommutes with the associated $R$-matrix. However, in said article we were not able to write such a coproduct beyond the short representation of this algebra. Here we will construct a coproduct map for the boost operator\footnote{By definition, such a map will be independent of the choice of a  representation.}. As in previous sections, we will only consider the boosts $\alg{J}_A$, as the action of any additional boost operator of the type discussed at the end of subsection \ref{subaut} can be obtained from Jacobi identities using the outer automorphism generators. A complete study of their formal inclusion in the coalgebra would require knowing the exact coproducts of the outer automorphism generators, which we plan to study in future publications.

We will argue that it is impossible to construct such a map using only the elements of the algebra and that we need to make use of elements of the $GL(2)^2$ outer automorphisms of $\mathfrak{su}(1|1)^2$. Interestingly, something similar was observed in the study of the universal $R$-matrix of the centrally extended $\mathfrak{psu}(2|2)$ algebra of the $AdS_5$ scattering problem \cite{Hecht1,Hecht2}. There it also was found that it is unavoidable to utilise the generators associated to the outer automorphisms, which are dual in the sense of the Killing form to the generators of the central extensions. The issue of the universal $R$-matrix encompassing the central extension has not been fully studied in the context of the $\mathfrak{su}(1|1)^2$ algebra (for a partial analysis, see Appendix C of \cite{MassiveT4}).

\subsection{The braided energy coproduct case}

Let us take the case of the left sector and the coproduct
\begin{equation}
	\Delta \alg{S}_L= \alg{S}_L \otimes e^{i p_L/4} + e^{-i p_L/4} \otimes \alg{S}_L \ , \qquad \Delta \alg{Q}_L= \alg{Q}_L \otimes e^{i p_L/4} + e^{-i p_L/4} \otimes \alg{Q}_L \ ,
\end{equation}
with similar coproduct for the right fermionic generators with $p_L$ substituted by $\pm p_R$. It is important to stress that this choice of coproducts forces us to set the central elements to have the following dependence on the momentum
\begin{align}
	\alg{H}_L&\propto \bigg\vert e^{i p_L/2} - e^{-i p_L/2}\bigg\vert \alg{1} \, \qquad \alg{H}_R\propto \bigg\vert e^{i p_R/2} - e^{-i p_R/2} \bigg\vert\alg{1} \ , \notag \\
	\alg{P} &\propto \alg{K} \propto \big(e^{i (p_L\pm p_R)/4} - e^{-i (p_L\pm p_R)/4} \big) \alg{1} \label{PKfromcoproduct}\ ,
\end{align}
where the $\pm$ in (\ref{PKfromcoproduct}) corresponds to the sign of the R-handed coproduct choice, not the choice of $\mathfrak{d}_{AB}$. These constraints are due to the fact that central generators of the algebra have to be co-commutative.

We should first comment on the issues this coproduct presents in the different algebras and how to deal with them. First of all, the algebra with $\mathfrak{d}_{AB}=0$ needs the central elements $\mathfrak{P}$ and $\mathfrak{K}$ to be separable to be consistent while the coproduct imposes them to be a trigonometric function of the sum (or difference) of the two momenta. The simplest way to deal with this issue is to set them to zero and consider the algebra $\mathfrak{d}_{AB}=0$ as two completely independent $\alg{su} (1|1)$ algebras. In addition, there is an apparent issue with the cases where $p_L \pm p_R=0$, but this is just a consequence of $\Delta \alg{P}$ and $\Delta \alg{K}$ becoming trivial in such cases, therefore eliminating the restriction that allowed to fix them. We will address this point in depth later.



In order to construct the coproduct map, we should start by looking at the form it takes when it is evaluated in a tensor product of short representations in the case $\mathfrak{d}_{AB}=1$ (i.e., when $\alg{S}_R\equiv \alg{Q}_L$ and $\alg{S}_L \equiv \alg{Q}_R$). This was constructed in \cite{Creature}\footnote{Equations (3.8) and (3.9) in \cite{Creature} possess typos, here we present the corrected expressions.} and is given by\footnote{A different expression for this map was proposed in \cite{JoakimAle}. Although it has different quasi-triangularity properties, its expression is similar enough so that the computation we are going to perform can be applied to it with minimal changes.}
\begin{equation}
	\Delta \alg{J}=\Delta_0 \alg{J} + \frac{e^{-\frac{i}{4} p} \otimes e^{\frac{i}{4} p}}{4} \left[ \alg{S} \otimes \alg{Q} + \alg{Q} \otimes \alg{S} \right] \ , \label{coproductJshort}
\end{equation}
where $\Delta_0 \alg{J}=\alg{J}_A \otimes \cos \frac{p}{2} + \cos \frac{p}{2} \otimes  \alg{J}_A$.

Although such map was constructed specifically to be an algebra homomorphism when considering the tensor product of short representations, it is easy to see that it is not an algebra homomorphism for the full algebra. The main obstruction is the appearance of factors of the form $\alg{Q}_L - \alg{S}_R$ and $\alg{Q}_R - \alg{S}_L$ that vanish for the short representation. One way to rephrase this problem is that we would like to find a way for the commutation relation
\begin{equation}
	[\left(  \alg{S}_L \otimes \alg{Q}_L + \alg{Q}_L \otimes \alg{S}_L \right), \Delta \alg{Q}_R]= e^{-\frac{i}{4} p_R} \alg{S}_L \otimes \alg{P} + \alg{P} \otimes e^{\frac{i}{4} p_R} \alg{S}_L \ 
\end{equation}
to become vanishing.

These unwanted terms pose a problem because all of them involve only the fermionic generator $\alg{S}_L$ while we started with $\alg{Q}_R$. There does not exist any generator in the algebra $\alg{su}(1|1)^2$ that can transform either of these two fermionic generators into the other, so we need to make use of the outer automorphisms to get rid of the unwanted terms. In particular, the combination of generators we are looking for is
\begin{equation}
	\alg{FT}_L=\alg{S}_L \otimes \alg{Q}_L + \alg{Q}_L \otimes \alg{S}_L -\alpha_R \left( \alg{P} \otimes t^\rho_+ + \alg{K} \otimes t^\lambda_+ \right) -\beta_R \left(t^\rho_+ \otimes \alg{P} + t^\lambda_+ \otimes \alg{K} \right) \ ,
\end{equation}
where $\alpha_A= -e^{\frac{i}{4} p_A} \otimes e^{\frac{i}{4} p_A}$ and $\beta_A= e^{-\frac{i}{4} p_A} \otimes e^{-\frac{i}{4} p_A}$. We can check that now the commutators with all the labelled $R$ fermionic generators vanish while all the new contributions vanish when applied to a labelled $L$ fermionic generator. A similar combination can be written for the right copy of the algebra
\begin{equation}
  \alg{FT}_R=\alg{S}_R \otimes \alg{Q}_R + \alg{Q}_R \otimes \alg{S}_R -\alpha_L \left( \alg{P} \otimes t^\lambda_- + \alg{K} \otimes t^\rho_- \right) -\beta_L \left(t^A_- \otimes \alg{P} + t^\rho_- \otimes \alg{K} \right) \ ,
\end{equation}
which commutes with the two labelled $L$ fermionic generators. These two linear combinations of generators are the cornerstone to tackle the problem of the coproduct for the $\mathfrak{d}_{AB} = \pm 1$ case.

\subsubsection{$\mathfrak{d}_{AB} = 0$ case}

We will start by focusing on the $\mathfrak{d}_{LR}=\mathfrak{d}_{RL}=0$ case. This is the simplest to treat, as the restriction $\alg{P}=\alg{K}=0$ completely separates the two algebras, so we can just consider equation~(\ref{coproductJshort}) and write the appropriate sub-indices. However, it is interesting for later arguments to upgrade the fermionic tail with the expressions we wrote above even though the new terms give no contribution
\begin{align}
\label{precisely}
\begin{split}
	\Delta \alg{J}_L (\mathfrak{d}_{AB} = 0) = \Delta_0 \alg{J}_L (\mathfrak{d}_{AB} = 0)+ \frac{e^{- i p/4} \otimes e^{i p/4}}{4} \alg{FT}_L  \ , \\
	\Delta \alg{J}_R (\mathfrak{d}_{AB} = 0) = \Delta_0 \alg{J}_R (\mathfrak{d}_{AB} = 0)+ \frac{e^{- i p/4} \otimes e^{i p/4}}{4} \alg{FT}_R \ .
\end{split}
\end{align}

\subsubsection{$\mathfrak{d}_{AB} = +1$ case}

Let us now focus on the case $\mathfrak{d}_{LR}=\mathfrak{d}_{RL}=1$. If we choose to set $\alg{J}_L=\alg{J}_R = \alg{J}$ with $\mathfrak{d}_{LR}=\mathfrak{d}_{RL}=1$, then the action of $\alg{J}$ is perfectly analogous to what the action of $\alg{J}_L+\alg{J}_R$ is in the case where $\mathfrak{d}_{LR}=\mathfrak{d}_{RL}=0$. At the level of the algebra, making this connection is unproblematic. At the level of the Hopf algebra, it is not. As we explained above, the $\mathfrak{d}_{LR}=\mathfrak{d}_{RL}=0$ case exhibits problems with coproducts and central elements, so we set the latter to zero. However, if we want to draw some parallels between how the boost operators act for the different values of $\mathfrak{d}_{AB}$, the upgraded tails we used in equation (\ref{precisely}) are necessary. In this way, they have the properties they need for us to write a similar relation for the coproduct of the boosts. Though, one needs to be be aware that by simply adding two $\mathfrak{d}_{LR}=\mathfrak{d}_{RL}=0$ coproducts, the derivative term would appear two-fold, hence it is necessary to subtract the so-obtained result by one $\Delta_0 \alg{J}$. This happens because at the end we have to identify $p_L=p_R$, so we are adding the derivative factor twice instead of once. Thus, the  combination  of  generators  that makes the coproduct map a homomorphism and that reduces to the correct expression for the short representations can be expressed as
\begin{gather}
\label{pre1}
	\Delta \alg{J} (\mathfrak{d}_{AB} = 1) = \Delta \alg{J}_L (\mathfrak{d}_{AB} = 0) +\Delta \alg{J}_R (\mathfrak{d}_{AB} = 0) -\Delta_0 \alg{J} \ , \\
	\Delta \alg{J} = \Delta_{0} \alg{J} +\alg{FT}_L +\alg{FT}_R =\Delta_0 \alg{J} + \frac{e^{-\frac{i}{4} p} \otimes e^{\frac{i}{4} p}}{4} \left\{  \alg{S}_L \otimes \alg{Q}_L + \alg{Q}_L \otimes \alg{S}_L +\alg{S}_R \otimes \alg{Q}_R + \alg{Q}_R \otimes \alg{S}_R \right. \notag \\
	\left. -\alpha \left[ \alg{P} \otimes (t^\rho_+ +t^\lambda_-) + \alg{K} \otimes (t^\lambda_+ + t^\rho_-) \right] -\beta \left[ (t^\rho_+ +t^\lambda_-)  \otimes \alg{P} + (t^\lambda_+ +t^\rho_-) \otimes \alg{K} \right] \right\} \ .
\end{gather}

\medskip

{\it $\bullet$ Short Representation}

\medskip

A consistency check of our coproduct is to study the representation where $\alg{S}_R=\alg{Q}_L=\alg{Q}$ and $\alg{Q}_R= \alg{S}_L=\alg{S}$, where it should reduce to the $\mathfrak{su}(1|1)$ boost-symmetry computed in \cite{Creature}, which we reproduce in equation~(\ref{coproductJshort}). One obstruction to evaluate $\Delta \alg{J}$ in this representation is the fact that the generators of the outer automorphism are not well-defined in this representation. However, we will see that they appear in a particular linear combination that is well-defined. In this representation we also have $\alg{H}_L=\alg{H}_R=\alg{P}=\alg{K}=\alg{H}$ and the coproduct of the boost becomes
\begin{equation}
	\Delta \alg{J}=\Delta_0 \alg{J} + \frac{e^{-\frac{i}{4} p} \otimes e^{\frac{i}{4} p}}{4} \left[ 2\alg{S} \otimes \alg{Q} + 2\alg{Q} \otimes \alg{S} -\alpha \alg{H} \otimes \alg{T} - \beta \alg{T} \otimes \alg{H} \right] \ ,
\end{equation}
where $\alg{T}= t^\lambda_+ + t^\lambda_- + t^\rho_+ + t^\rho_-$. For this representation we can see that $[\alg{T} , \alg{Q}]=\alg{Q}$ and $[\alg{T} , \alg{S}]=\alg{S}$, which implies that
\begin{equation}
	\left[ 2\alg{S} \otimes \alg{Q} + 2\alg{Q} \otimes \alg{S} -\alpha \alg{H} \otimes \alg{T} - \beta \alg{T} \otimes \alg{H} , \Delta X \right]=\left[ \alg{S} \otimes \alg{Q} + \alg{Q} \otimes \alg{S} , \Delta X \right] \ ,
\end{equation}
for $X=\alg{Q}$ or $\alg{S}$. Thus our coproduct map for the boost for the $\mathfrak{d}_{LR}=\mathfrak{d}_{RL}=1$ reduces to the appropriate one in this representation.

\subsubsection{$\mathfrak{d}_{AB} = -1$ case}

Let us now move to the case $\mathfrak{d}_{LR}=\mathfrak{d}_{RL}=- 1$. First of all, we find that equation~(\ref{PKfromcoproduct}) seems to fix two of the central elements to be trivial in this case, due to the relation between the momenta. This is not the case because, although the boost behaves in this case very similarly to the case $\mathfrak{d}_{LR}=\mathfrak{d}_{RL}=1$, the coproduct structure allows us more freedom as the central elements $\alg{P}$ and $\alg{K}$ fulfil now
\begin{align*}
	\Delta \alg{P} &= \alg{P} \otimes 1 + 1 \otimes \alg{P} \ , & \Delta \alg{K} &= \alg{K} \otimes 1 + 1 \otimes \alg{K} \ .
\end{align*}
Thus any dependence on the momentum now satisfies the cocommutativity property and the behaviour for them we deduced in equation~(\ref{PKfromcoproduct}) is not valid any more. Nevertheless, there is nothing special about this point and from the commutation relations between $\alg{P}$, $\alg{K}$ and $\alg{J}$, we can fix the four central elements to be
\begin{equation}
	\alg{H}_L \propto \alg{H}_R\propto \alg{P} \propto \alg{K} \propto (e^{i p/2} - e^{-i p/2}) \alg{1} \ ,
\end{equation}
and a similar construction to the case $\mathfrak{d}_{LR}=\mathfrak{d}_{RL}=1$ follows.

Finally, we want to comment that choosing the opposite sign for $p_R$ in (\ref{PKfromcoproduct}) just exchanges the behaviour between the cases $\mathfrak{d}_{LR}=\mathfrak{d}_{RL}=\pm 1$, so the same arguments apply there.

%

\subsection{The unbraided energy coproduct case}

 In the above subsection, we constructed the coproduct of the boost generator for the case we denoted as ``bosonically braided coproduct'' in \cite{Creature}. In this subsection, we want to complement it by computing the coproduct of the boost in the case we denoted as ``bosonically unbraided'' in said article.

This bosonically unbraided coproduct is defined by the following choice of coproducts for the fermionic generators
\begin{align*}
\Delta \alg{Q}_A &= \alg{Q}_A \otimes e^{i \frac{p}{4}} + e^{-i \frac{p}{4}} \otimes \alg{Q}_A \\
\Delta \alg{S}_A &= \alg{S}_A \otimes e^{-i \frac{p}{4}} + e^{i \frac{p}{4}} \otimes \alg{S}_A ,
\end{align*}
In contrast with the bosonically braided case, here the coproduct of the boost also involves the hypercharge operator. In particular, for the short representation we found that 
\begin{gather}
\Delta \alg{J}_L= A (p_1, p_2) \alg{J}_L \otimes 1 + B (p_1, p_2) 1 \otimes \alg{J}_L + F_+(p_1, p_2) \alg{S} \otimes \alg{Q} \nonumber \\ 
\qquad + F_-(p_1 , p_2) \alg{Q} \otimes \alg{S} + G(p_1 , p_2) \left[ \alg{B} \otimes \alg{1} - \alg{1} \otimes \alg{B} \right] \ .
\end{gather}
There are multiple subtleties and ambiguities in the process of fixing the functions $A$, $B$, $F_+$, $F_-$ and $G$, which are discussed at length in \cite{Creature}. One of the possible ways of fixing them is
\begin{align*}
	A (p_1, p_2) &=B(p_2 , p_1)=\cot \left(\frac{p_1}{2}\right) \cot \left(\frac{p_2 -p_1}{2}\right) \\
	F_\pm(p_1, p_2) &= e^{\pm i\frac{p_1+p_2}{4}} \cot \left(\frac{p_1 -p_2}{2}\right) \left[ i\csc \left(\frac{p_1}{2}\right) \csc \left(\frac{p_2}{2}\right) - i \pm  \cot \left(\frac{p_1 +p_2}{2}\right) \right]  \ , \\ G(p_1 , p_2) &= -\frac{1}{16} i h \cos \left(\frac{p_1-p_2}{4}\right) \csc \left(\frac{p_1+p_2}{4}\right)\ .
\end{align*}

In order to construct an analogous $\alg{FT}_A$ for this case, we also need to substitute any $\alg{B}$ operator present in $\Delta \alg{J}_A$ by another operator that acts as the usual hypercharge on $A$-handed operators, whereas they would have to vanish on operators with the opposite handedness. We call these operators $\alg{B}_A$ and they can be written in terms of the outer automorphisms as
\begin{align*}
- i \alg{B}_R &= t^\lambda_0 - t^\rho_0 - t^\lambda_3 - t^\rho_3 \, ,\\
- i \alg{B}_L &= t^\lambda_0 - t^\rho_0 + t^\lambda_3 + t^\rho_3 \, .
\end{align*}
With those operators we can write the following upgraded version of the tail
\begin{align}
\alg{FT}_L&= G \left[ \alg{B}_L \otimes \alg{1} - \alg{1} \otimes \alg{B}_L \right] + F_+  \left[ \alg{S}_L \otimes \alg{Q}_L -\beta_R t^\rho_+ \otimes \alg{P} +\beta_R \alg{K} \otimes t^\lambda_+ \right] \notag \\
&+ F_- \left[ \alg{Q}_L \otimes \alg{S}_L -\alpha_R \alg{P} \otimes t^\rho_+ + \alpha_R t^\lambda_+ \otimes \alg{K} \right] \ , \\
\alg{FT}_R&= G \left[ \alg{B}_R \otimes \alg{1} - \alg{1} \otimes \alg{B}_R \right] + F_+  \left[ \alg{Q}_R \otimes \alg{S}_R +\alpha_L t^\rho_- \otimes \alg{K} -\alpha_L \alg{P} \otimes t^\lambda_- \right] \notag \\
&+ F_- \left[ \alg{S}_R \otimes \alg{Q}_R +\beta_L \alg{K} \otimes t^\rho_- - \beta_L t^\lambda_- \otimes \alg{P} \right] \ .
\end{align}
Here, we have again  $\alpha_A= -e^{\frac{i}{4} p_A} \otimes e^{\frac{i}{4} p_A}$ and $\beta_A= e^{-\frac{i}{4} p_A} \otimes e^{-\frac{i}{4} p_A}$.

Having constructed the $\alg{FT}_A$ for this choice of fermionic coproducts, we can follow the prescription we used above, that is
\begin{align*}
	\Delta \alg{J}_L (\mathfrak{d}_{AB} = 0) &= \Delta_0 \alg{J}_L (\mathfrak{d}_{AB} = 0)+ \alg{FT}_L  \ , \\
	\Delta \alg{J}_R (\mathfrak{d}_{AB} = 0) &= \Delta_0 \alg{J}_R (\mathfrak{d}_{AB} = 0)+ \alg{FT}_R \ , \\
	\Delta \alg{J} (\mathfrak{d}_{AB} = 1) &= \Delta \alg{J}_L (\mathfrak{d}_{AB} = 0) +\Delta \alg{J}_R (\mathfrak{d}_{AB} = 0) -\Delta_0 \alg{J} = \Delta_{0} \alg{J} + G \left( \alg{B} \otimes \alg{1} - \alg{1} \otimes \alg{B} \right) \\
	&+ F_+ \left( \alg{S}_L \otimes \alg{Q}_L + \alg{Q}_R \otimes \alg{S}_R - \beta \left( t^\rho_+ \otimes \alg{P} - \alg{K} \otimes t^\lambda_+ \right) - \alpha \left( \alg{P} \otimes t^\lambda_- - t^\rho_- \otimes \alg{K} \right) \right)
 \\
& + F_- \left( \alg{Q}_L \otimes \alg{S}_L + \alg{S}_R \otimes \alg{Q}_R - \alpha \left( \alg{P} \otimes t^\rho_+ - t^\lambda_+ \otimes \alg{K} \right) - \beta \left( t^\lambda_- \otimes \alg{P} - \alg{K} \otimes t^\rho_- \right) \right) \, ,
\end{align*}
keeping in mind that we identify $p_L = p_R \equiv p$ and that  $\alg{B}_R + \alg{B}_L = \alg{B}$. \par

\section{Conclusions}

In this article, we have studied different consistent extensions of the centrally extended $\alg{su}(1|1)^2$ algebra to the case where each copy depends on a different momenta. This shed some light on how to construct the coproduct map for the boost operator we can add to such an algebra, generalising the results obtained in \cite{Creature} for short representations to any representation.

\begin{figure}[h]
\begin{align*}
  \begin{tikzpicture}
    \node (A) at (0,0) {$\mathfrak{H}_L \mathfrak{d}_{LR} = \zeta \mathfrak{H}_R$};
    \node (B) at (-4,1.8) {$\mathfrak{d}_{AB} = +1$};
    \node (C) at (-4,-1.8) {$\mathfrak{d}_{AB} = -1$};
    \node (D) at (4.1,0) {$\mathfrak{d}_{AB} = 0$};
    \node (E) at (4.3,2) {$\substack{\mathfrak{d}_{LR} = 0\\ \mathfrak{d}_{RL} = \zeta \mathfrak{H}_L}$};
    \node (F) at (4.3,-2) {$\substack{\mathfrak{d}_{RL} = 0 \\ \mathfrak{d}_{LR} = \zeta \mathfrak{H}_R}$};
    \node (L1) at (-6,-4) {\textit{\textbf{differential algebras}}};
    \node (L2) at (5.5,-4) {\textit{\textbf{separable algebras}}};
    \begin{scope}[fill opacity=0.4]
\draw (2.5,0) circle (3.8cm);
\draw (-2.5,0) circle (3.8cm);
      \end{scope}
    \draw[thick,<->] (C) -- (B) node[midway,sloped,left,rotate=270] {$p_L = \pm p_R$};
    \draw[thick,->] (B) -- (A) node[midway,sloped,above left,rotate=0] {(\ref{twoH})};
    \draw[thick,->] (C) -- (A) node[midway,sloped,below left,rotate=0] {(\ref{twoH})};
    \draw[thick,->] (D) -- (A) node[midway,sloped,above,rotate=0] {(\ref{reex}), (\ref{twoH})};
    \draw[thick,->] (D) -- (E) node[midway,sloped,above left,rotate=278] {(\ref{arr1})};
    \draw[thick,->] (D) -- (F) node[midway,sloped,above right,rotate=82] {(\ref{arr2})};
  \end{tikzpicture}
\end{align*}
\caption{On the left, we can see how the differentiable algebras are related to each other, whereas on the right we can see the genealogy of separable algebras. Notice the curious relation of $\mathfrak{H}_L \mathfrak{d}_{LR} = \zeta \mathfrak{H}_R$ with both algebra types.}\label{figure}
\end{figure}
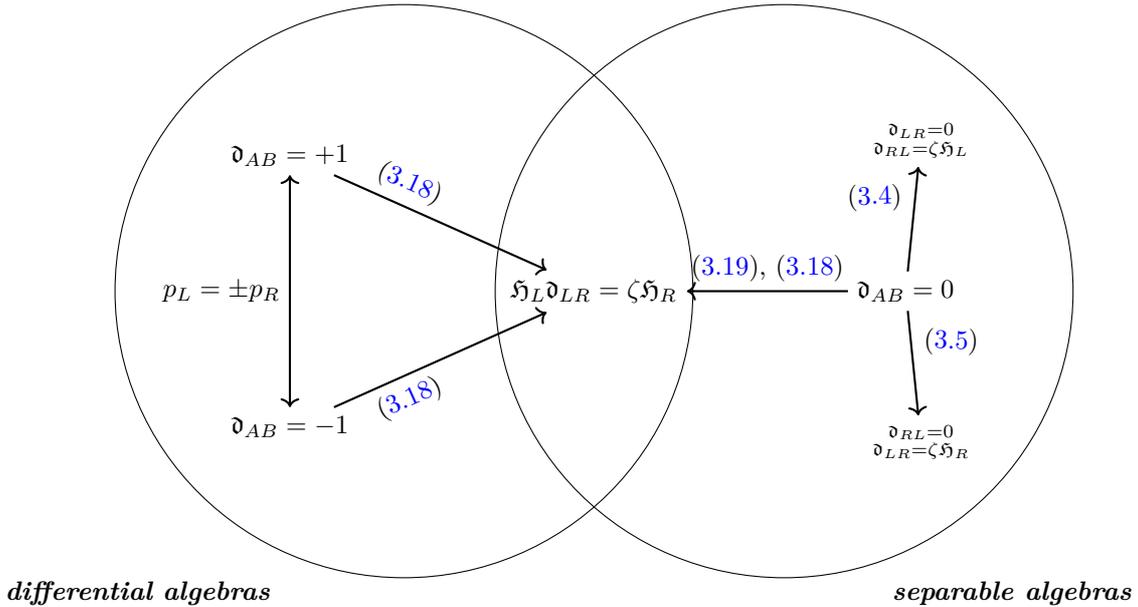

First, we investigated possible ways the boost operator is able to adjointly act on generators of opposite handedness (and by extension, also the central elements). By imposing consistency of the Jacobi identities, we  arrive at six different algebras, some of which are the handedness-transformed of each other. Essentially, these can be classified in terms of two categories, as schematically depicted in Figure~\ref{figure}.


These names are explained by the most characteristic property of each of the two sets of algebras. The three separable algebras are characterised by the fact that they can be split into two well separated subalgebras that do not talk to each other. The differential algebras are characterised by the equation $\mathfrak{d}_{LR} \mathfrak{d}_{RL}=1$, which can be understood as a consequence of the inverse function theorem if we understand the boost as a differential operator. The $\mathfrak{H}_L \mathfrak{d}_{LR} = \zeta \mathfrak{H}_R$ is interesting, as it can both be understood as differential and separable at the same time.

After this close study of different algebras, we were able to construct coproduct maps for the boost operator $\alg{J}_A$ for each of the cases. The crucial point in our construction is that such coproducts cannot be built with only elements of our algebra and we need to make use of the generators associated to the outer automorphisms group of $\alg{su}(1|1)^2$ \cite{B6,Hecht1,Hecht2}. They appear in the tail of $\Delta \alg{J}_A$ as key ingredients to cancel out unwanted fermionic contributions in commutators. We have shown that it is enough for us to compute $\Delta \alg{J}_A$ for the case $\mathfrak{d}_{AB} = 0$, as the remaining cases can be constructed by appropriate linear combinations of the fermionic tails associated to those.

The most immediate extension of this work is to consider the massive case, as we have restricted ourselves to the massless one so far. Another task would be to construct the coproduct map associated to the symmetry which in $AdS_5 \times S^5$ was computed for short representations by \cite{BorAle}. As in the case studied here, we expect the outer automorphisms to play a central rôle also in those situations, as they are known to be dual (in the sense of Hopf algebra) to the generators associated to the  central extension \cite{Hecht1,Hecht2}. In addition, it would be interesting to repeat the construction of the dual of the algebra performed in those two articles in our setting, as this will confirm that central generators and outer automorphisms are dual to each other. Furthermore, we would be able to check if the $R$-matrix obtained through that procedure still quasi-cocommutes with the coproduct of the boost we computed here.

An interesting point we have not studied in this paper is the antipode map and its action on the newly constructed coproduct map. An antipode would extend the bi-algebraic structure we have described in this article to a Hopf algebra. However, this computation would require a more profound knowledge of the action of the coproduct and the counit on the outer automorphisms that is beyond the scope of this article. Nevertheless, we plan to address this point in future publications.

\section*{Acknowledgements}
We are grateful to Vidas Regelskis for reading the manuscript and providing very useful comments. LW is funded by a University of Surrey Doctoral College Studentship Award. This work is supported by the EPSRC-SFI grant EP/S020888/1 {\it Solving Spins and Strings}. We also thank the anonymous referee for his or her helpful input that has led to an improvement of the manuscript.

No data beyond those presented and cited in this work are needed to validate this study.

\end{document}